\documentclass[prl,
	twocolumn,
	showpacs,superscriptaddress]{revtex4}

\usepackage{amssymb,amsmath,bm,graphicx,amsthm}

\newcommand{\ket}[1]{|#1\rangle}
\newcommand{\bra}[1]{\langle #1|}
\newcommand{\tr}{\mathrm{tr}}
\newtheorem*{definition}{Definition}

\begin{document}

\title{Strings, Projected Entangled Pair States, 
	and variational Monte Carlo methods}

\author{Norbert Schuch}
\affiliation{Max-Planck-Institut f\"ur Quantenoptik, 
    Hans-Kopfermann-Str.\ 1, D-85748 Garching, Germany.}
\author{Michael M.\ Wolf}
\affiliation{Max-Planck-Institut f\"ur Quantenoptik, 
    Hans-Kopfermann-Str.\ 1, D-85748 Garching, Germany.}
\author{Frank Verstraete}
\affiliation{Fakult\"at f\"ur Physik, Universit\"at Wien, 
Boltzmanngasse 5, A-1090 Wien, Austria.} 
\author{J.\ Ignacio Cirac}
\affiliation{Max-Planck-Institut f\"ur Quantenoptik, 
    Hans-Kopfermann-Str.\ 1, D-85748 Garching, Germany.}

\begin{abstract}
We introduce string-bond states, a class of states obtained by placing strings
of operators on a lattice, which encompasses the relevant states in
Quantum Information.  For string-bond states, expectation values of local
observables can be computed efficiently using Monte Carlo sampling, making
them suitable for a variational abgorithm which extends DMRG to higher
dimensional and irregular systems.  Numerical results demonstrate the
applicability of these states to the simulation of many-body sytems.
\end{abstract}


\maketitle

\textit{Introduction.---}%
Explaining the properties of quantum many-body systems is a central topic
in modern physics.  Its difficulty is closely related to the hardness of
finding a practical description of the quantum state of those systems.
Therefore, results are usually derived by using either analytic
approximations or numerical methods, as Quantum Monte Carlo
(QMC)~\cite{qmc} or the Density Matrix Renormalization Group
(DMRG)~\cite{dmrg}.  While DMRG works extremely well for one-dimensional
systems, Monte Carlo proved very successful also in describing the
behaviour of non-frustrated systems in higher dimensions.  Recently, an
extension of DMRG to two-dimensional systems has been developed which
gives good results even for frustrated systems and time
evolution~\cite{vidal:timeevol,peps-alg,peps-alg-data}.  It is based on
Projected Entangled Pair States (PEPS) which describe Nature at low
temperatures very well as has been proven by
Hastings~\cite{hastings-mpdo,hastings-mps}.  However, the class of states
is too large~\cite{cplx-peps}, leading to an unfavourable scaling of the
method in more than two dimensions or for periodic boundary conditions
(PBC).  Moreover, the algorithm relies on the underlying lattice structure
so that irregular systems cannot be handled in a simple way.  A subclass
of PEPS which can be dealt with in an efficient way while keeping the
power of the full family may be a solution to these issues.

In this paper, we intruduce a new class of states called \emph{string-bond
states}.  String-bond states form a subclass of PEPS which contains the
relevant states in Quantum Information, as e.g.\ the toric code or
the cluster state.  Since expectation values can be computed easily
on string-bond states using Monte Carlo sampling, they can be used to
build a variational Monte Carlo algorithm for finding ground states.  The
central idea is to create the states by placing strings of operators on a
lattice, which naturally extends the class of Matrix Product States (MPS)
underlying DMRG to arbitrary geometries. Thus, the method combines the
strengths of DMRG/PEPS and Monte Carlo: It can be applied to
three-dimensional systems, systems with periodic boundary conditions, or
general geometries by adapting the string pattern, but also
works for frustrated or fermionic systems which cannot be dealt with using
Monte Carlo.
 At the same time, the computational
resources scale favourably in the relevant parameters. We have implemented
the method and successfully demonstrated its applicability.

\textit{String-bond states.---}%
Consider a classical spin system with configurations
$n=(n_1,\dots,
n_N)\in \{1,\dots,d\}^N$ equipped with a probability distribution $p(n)$,
and an efficiently
computable function $f(n)$.   The expectation value of $f(n)$, 
$\sum_n p(n) f(n)$, can be computed using Monte Carlo---this is, by
randomly sampling $f(n)$ according to the distribution $p(n)$---whenever 
$p(n)$ can be computed efficiently up to normalization. 
Turning towards quantum systems, for a state $\ket{\psi}$ and an
observable $O$
\begin{equation}
\bra{\psi}O\ket{\psi}=
    \sum_n \langle\psi\ket{n}\bra{n}O\ket{\psi}
=\sum_n p(n) \frac{\bra{n}O\ket{\psi}}{\bra{n}\psi\rangle}
\label{eq:var-mc}
\end{equation}
with $p(n)=|\langle n\ket\psi|^2$, and therefore $\bra\psi O\ket\psi$ can
be evaluated using Monte Carlo whenever $\langle n\ket\psi$ and $\bra n
O\ket\psi$ can be computed efficiently. The latter reduces to
$\langle \tilde n\ket\psi$ whenever $O=\sum D_kP_k$ with $D_k$ diagonal
and $P_k$ permutations. In particular, this holds for 
local $O$ (local meaning small support, as e.g.\
two-point correlations) and products of Paulis, as  e.g.\ string order
parameters.

\begin{figure}[t]
\includegraphics[width=0.95\columnwidth]{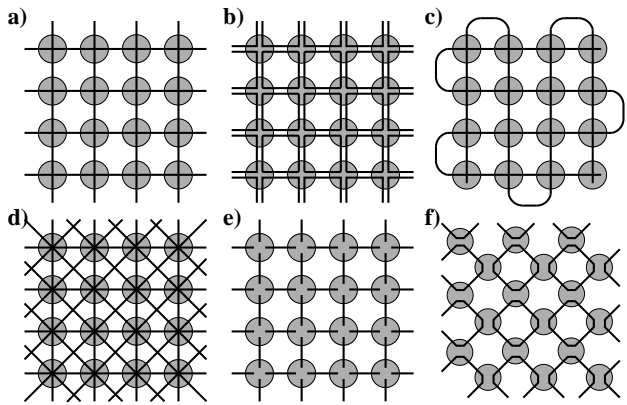}
\caption{\label{fig:string-arrangements}
Various string patterns. In a), c) and d), the string are long lines -- in
particular, c) directly generalizes the DMRG ansatz, and since d)
contains a), it gives a larger class of states.  In b)
and f), the strings form small loops; pattern f) 
underlies the toric
code state.  In e), the strings have length one, which suffices
e.g.\ for the cluster state or the coherent version of classical
Gibbs states.  Patterns a), b), and both atop of each other have been
implemented numerically.  }
\end{figure}

To build a variational Monte Carlo method, one therefore has to construct
states for which $\langle n\ket\psi$ can be computed
efficiently. One such class is given by 
Matrix Product States (MPS)~\cite{mps-general}, the class of states
underlying DMRG. An MPS with \emph{bond dimension} $D$ has the form
\begin{equation}
\ket\psi=\sum_{n_1,\dots,n_N} 
\tr\left[M^1_{n_1}\cdots M^n_{n_N}\right]\ket{n_1,\dots,n_N}\ 
\label{eq:mps}
\end{equation}
where each $M^i_{n_i}$ is a $D\times D$ matrix, so that
$\langle n\ket\psi$ is given by the trace which 
can be computed efficiently.  We generalize this to arbitrary geometries
as follows.
\begin{definition}
A state of $N$ $d$-level spins is a \emph{string bond state} if there exists a
local basis $\ket{n}=\ket{n_1}\cdots\ket{n_N}$ and a set of 
strings $s\in\mathcal S$ (i.e.\ $s$ is an  ordered subset of
$\{1,\dots,N\}$) such that
\begin{equation}
\label{eq:stringstate}
\langle n\ket\psi=\prod_{s\in\mathcal S}
    \tr\left[\prod_{x\in s}M^{s,x}_{n_x}\right]
\end{equation}
for some complex $D\times D$ matrices $M^{s,n}_{n_x}$. 
Here, the product over $x\in s$ is over the sites $x$ in the order 
in which they appear in the string.
\end{definition}
Note that the trace of a product of operators in (\ref{eq:stringstate})
resembles the structure of MPS (\ref{eq:mps}). Some possible string
arrangements are shown in Fig.~\ref{fig:string-arrangements}.

The key point in the definition is the factorization of $\langle
n\ket\psi$ into efficiently computable coefficients which we chose
to be the trace of a matrix product. However, there are many
more natural choices, as small ``blobs'' with e.g.\ a PEPS parametrization
or tree tensor networks~\cite{ttn}. A special
case is given by quantum states corresponding to thermal states of
classical models~\cite{mix-peps} which have strings between neighboring
sites only, cf.~Fig.~\ref{fig:string-arrangements}e.

A variational ansatz based on string-bond states generalizes
DMRG beyond one-dimensional systems by combining it with Monte Carlo
methods. In particular, due to the flexibility in the arrangement of the
strings it can be adapted to arbitrary geometries, and the accuracy can be
increased either by increasing $D$ or by adding more strings.  Clearly,
the factorization of (\ref{eq:stringstate}) does not imply that the
string-bond states themselves factorizes into Matrix Product States, and 
in fact they contain a large variety of relevant states.

\textit{Properties.---}%
Let us first clarify the relation between string-bond states and
Projected Entangled Pair States (PEPS)~\cite{peps-alg}.
 To define a PEPS on any
graph, place
maximally entangled bonds $\sum_{i=0}^{D-1}\ket{i}\ket{i}$ on each
edge---associating each virtual spin with one vertex---and
apply a linear map $P^{[v]}$ on each vertex $v$ which maps the
virtual spins to the $d$-dimensional physical spin at $v$.

\begin{figure}[t]
\includegraphics[width=0.95\columnwidth]{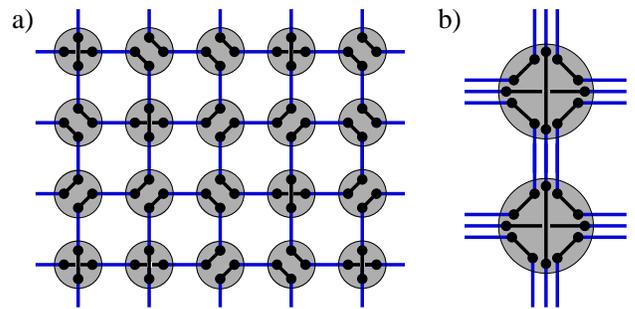}
\caption{\label{fig:peps}
a) A PEPS with factorizing projectors---connected by the maximally
entanged bonds (blue)---yields a string-bond state.  
b) To convert a
general string state with many strings (we illustrate patterns
Fig.~\ref{fig:string-arrangements}a and \ref{fig:string-arrangements}b
together) into a PEPS, each string is routed over a separate bond.  }
\end{figure}

For clarity, we restrict to a 2D lattice with periodic boundaries.
Consider a PEPS with linear maps
\begin{equation}
P^{[i,j]}=\sum_{s=0}^{d-1}\ket{s}\bra{\phi^{a,s}_{i,j}}
\bra{\phi^{b,s}_{i,j}}
\label{eq:string-proj} 
\end{equation}
at site $(i,j)$,
where $\bra{\phi^a}$ and $\bra{\phi^b}$ act on two virtual spins each: 
One readily sees that together with the bonds they form strings
(Fig.~\ref{fig:peps}a),
and $\langle n\ket\psi$ is given by the product of the
overlaps of all strings.   This generalizes the
states corresponding to classical thermal states for which $P$
factorizes completely~\cite{mix-peps}.
On the other hand, every string-bond state can be written as a PEPS, even
if at some edges many strings come to lie atop of each
other. In that case, one places several maximally entangled bonds on that
edge and uses one of them for each string. The product
over the strings results in a factorizing map $P$ as
in (\ref{eq:string-proj}), where the number of bipartite projectors
$\bra{\phi_{i,j}^{x,s}}$ 
equals the number of strings (see Fig.~\ref{fig:peps}b). 
Thus, the map pertains an efficient description, while the total bond
dimension scales exponentially in the number of strings.

String-bond states are complete, i.e.\ every state can be written as a
string-bond
state for large enough $D$, even of form (\ref{eq:string-proj}).  This is
easily seen by using one string which covers the whole system as in
Fig.~\ref{fig:string-arrangements}c and using
the completeness of MPS~\cite{mps-reps}. 

String-bond states encompass a variety of relevant states in Quantum
Information.  First, this holds for all MPS
as e.g.\ the GHZ or the W state~\cite{mps-reps}.  
Any (generalized) weighted graph state, as e.g.\ the cluster
state~\cite{gwgs}, and thus all stabilizer states~\cite{stab-graph}
are string-bond
states with $D=2$~\cite{cluster-peps}.  The same is true for Kitaev's
toric code
state~\cite{kitaev:toriccode},
using the pattern Fig.~\ref{fig:string-arrangements}f:
The strings from loops, and 
the weight of each classical configuration is the parity
of the four spins on the loop, corresponding
to $P= \ket{0}\bra{\psi^+}\bra{\psi^+}+\ket{1}\bra{\psi^-}\bra{\psi^-}$~%
\cite{mix-peps}.  This results in a superposition of all states with an
even number of $\ket{1}$'s on any loop, which is exactly the string
condensate which gives the toric code state. In a sense, the string-bond states
extend the construction of Kitaev~\cite{kitaev:toriccode} and
Levin and Wen~\cite{wen:stringnets} and may help to add new insight
into Topological Quantum Computation. 
Note that both the cluster and the toric code state have a block entropy
which scales as the area, and thus string states can achieve the 
entropic area
law.

\textit{Variational ansatz.---}%
The fact that expectation values of local observables and thus the 
energy of a local Hamiltonian $H$ can be computed efficiently on
string-bond states allows to use them as a variational ansatz for the
computation of ground state properties. Therefore, pick a string $s$,
a site $x$ on the string, and minimize the 
energy over the corresponding
matrices $(M^{s,x}_{n_x})_{n_x=1}^d\equiv A$ (where $A$ is a three-index
tensor). By  iterating this protocol until it converges, one gets a 
better and better approximation to the ground state.

For the optimization, we can use the linearity of string-bond states in 
$A$,
\begin{equation}
E(\psi_A)=\frac{\bra{\psi_A}H\ket{\psi_A}}{\bra{\psi_A}\psi_A\rangle}=:
    \frac{ \bm{\bra{A} X\ket A}}{\bm{\bra A Y\ket A}}\ ,
\label{eq:qform}
\end{equation}
where we have explicitly denoted the dependence of the string-bond state
$\ket{\psi_A}$ on $A$.
$\bm{\bra A X \ket A}$ denotes a 
quadratic form in $A$, i.e.\ $\bm{\ket A}$ is the vectorized
form of $A$, where we use boldface to avoid confusion with vectors in
states space.
Minimizing (\ref{eq:qform})
with respect to $A$ is a generalized eigenvalue problem
and can be solved efficiently.

However, there are a few issues which render this approach
infeasible. Firstly, $X$ and $Y$ have $(dD^2)^2$ degrees of freedom---for
$d=2$, $D=6$, this is over $5000$---and one would have to do the
corresponding number of Monte Carlo runs for a \emph{single} optimization
step.
Moreover, each run would contribute to the error of $X$ and
$Y$, thus requiring a high sampling accuracy. To overcome this problem,
we use a Monte Carlo technique called \emph{reweighting}. The idea is to 
replace the sampling over a distribution $p(n)$ by the sampling over
some related distribution $p_0(n)\approx p(n)$,
\begin{equation}
\frac{\sum_n p(n) f(n)}{\sum_n p(n)}
=\frac{\sum_n p_0(n) \frac{p(n)}{p_0(n)}f(n)}{
    \sum_n p_0(n)\frac{p(n)}{p_0(n)}}\
.
\label{eq:mc-rew}
\end{equation}
In our case, $p(n)=|\bra{n}\psi_A\rangle|^2$,
$p_0(n)=|\bra{n}\psi_{A_0}\rangle|^2$ 
(where $A_0$ denotes the initial value of $A$), and
$f(n)=\bra{n}H\ket{\psi_A}/{\bra n\psi_A\rangle}$,
cf.\ Eq.\ (\ref{eq:var-mc}). Now define $\bm{\ket{a_n}}$ and
$\bm{\ket{b_n}}$ via the linear functionals 
\begin{equation}
\bm{\langle a_n\ket A}=
\frac{\bra{n}H\ket{\psi_A}}{\bra{n}\psi_{A_0}\rangle}\ , \
\bm{\langle b_n\ket A}=
\frac{\bra{n}\psi_A\rangle}{\bra{n}\psi_{A_0}\rangle}\ .
\label{eq:an-bn}
\end{equation}
Then one can readily check using (\ref{eq:var-mc}) and (\ref{eq:mc-rew})
that the matrices $\bm X$ and $\bm Y$ in (\ref{eq:qform}) are
\begin{equation}
\bm X=\sum_n p_0(n) \bm{\ket{b_n}\bra{a_n}}\ , \ 
\bm Y=\sum_n p_0(n) \bm{\ket{b_n}\bra{b_n}}\ ,
\label{eq:sampleXY}
\end{equation}
i.e., we can compute $X$ and $Y$ with a single Monte Carlo run.

The second problem is the inaccuracy of $\bm X$ and $\bm Y$
due to the finite sampling length: In particular, errors in the kernel of
$\bm Y$ will very often lead to to a wrong minimum.
We overcome this problem by 
moving along the gradient of $E(\psi_A)$
by a small distance.
With (\ref{eq:qform}) and
$\bm{\bra{A_0}Y\ket{A_0}}=1$, we find that
\begin{equation}
\left.\mathrm{grad}_AE(\psi_{A})\right|_{A=A_0}=
    \bm{Y\ket{A_0}\bra{A_0}X\ket{A_0}-X\ket{A_0}}
\label{eq:gradient}
\end{equation}
which only depends on absolute errors.

At this stage, we have an applicable algorithm.  An extra speedup of
$(dD^2)$ is obtained by directly sampling the gradient:
From (\ref{eq:sampleXY}), (\ref{eq:gradient}) 
and $\bm{\langle b_n\ket{A_0}}=1$ [Eq.~(\ref{eq:an-bn})],
\[
\left.\mathrm{grad}_{A}E\right|_{A=A_0}=
\sum_n p(n)\bm{\ket{b_n}}\left[
    \left(\sum_m p(m)h_m\right)-h_n\right]\ ,
\]
where we have defined
\[
h_n:=\bm{\langle a_n\ket{A_0}}=
    \frac{\bra n H\ket{\psi_{A_0}}}{\bra n\psi_{A_0}\rangle}\ .
\]
Note that for local $H$, $h_n$ only depends on the strings which
intersect with $H$, and similarly $\bm{\ket{b_n}}$ only depends on the
string which contains $A$.
As $h_n$ is independent of the site to be optimized, one can compute
the gradients for \emph{all} sites from the same sample, 
and move along all of them simultaneously which gives another improvement
of the order of the lattice size.

\textit{Numerical results.---}%
In order to demonstrate the suitability of string-bond states for ground state
calculations, we have implemented a simple non-opimized Matlab program for
the string patterns Fig.~\ref{fig:string-arrangements}a (lines) and
Fig.~\ref{fig:string-arrangements}a+b
(lines+loops) on a 2D
lattice with PBC.  Adding loops typically leads to a significant
improvement as it gives full control of correlations also to the first
diagonal neighbor. Additional strings which supply connections to more
diagonal neighbors further increase the accuracy.

\begin{figure}[t]
\includegraphics[width=0.95\columnwidth]{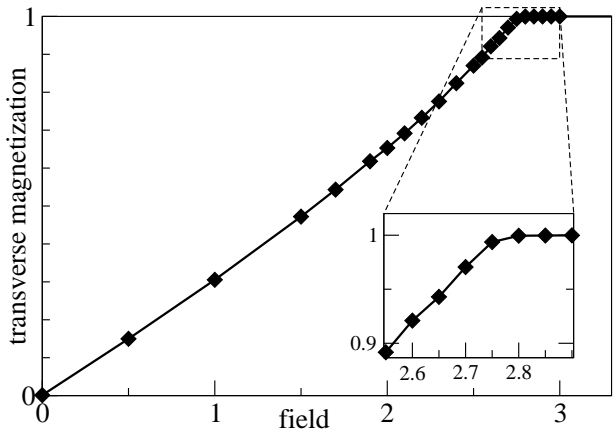}
\caption{\label{fig:frust-xx}
Magnetization for a frustrated $XX$ model (where each plaquette is
frustrated) as a function of the transverse field on a $10\times10$
lattice with PBC.
 Note that there is no other method available which can deal
with such systems.}
\end{figure}

We have tested our method by comparing it to the general PEPS
algorithm~\cite{peps-alg,vamu:private}, which is the only available
general benchmark for frustrated systems.  
For the frustrated $XX$ model on an $8\times 8$ 
lattice with open boundary conditions (OBC),
the general PEPS method gives  $E=-92.39$ for $D=4$, whereas string-bond states
with lines+loops give $E=-93.31\pm0.02$ with $D=8$, while at the same time
being about $30$ times faster.  Apparently, the
entanglement structure of frustrated systems is well reproduced only for
large $D$, suggesting that string-bond states are very suitable to describe
such systems due to their favourable scaling, and since they are not
restricted to OBC. Fig.~\ref{fig:frust-xx} shows the magnetization for a
frustrated $XX$ model as a function of the transverse field for a
$10\times10$ PBC lattice (computed with the lines+loops setup), 
for which we have no method to compare with.

\begin{figure}[t]
\includegraphics[width=0.95\columnwidth]{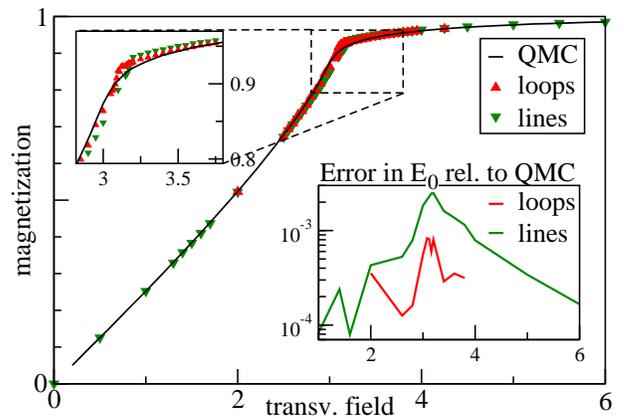}
\caption{\label{fig:ising}
Magnetization for the $10\times10$ PBC Ising model with transverse field:
results for Quantum Monte Carlo (QMC) and the string setups
Fig.~\ref{fig:string-arrangements}a (lines) and
Fig.~\ref{fig:string-arrangements}a+b (loops). The right inset shows the
relative error in the ground state energy compared to QMC
as a function of the field.}
\end{figure}

To compare the algorithm for PBC, we have therefore investigated the
 2D Ising model with transverse field 
and compared the results to Quantum Monte Carlo
(QMC)~\cite{tommaso:private}.  Fig.~\ref{fig:ising} shows the
magnetization and the relative error in energy compared to QMC as a
function of the field. Already the two basic string setups used reproduce
both energy and magnetization very well.

In all cases, we start with a very low number of sampling points,
$M=2000$, and with $D=2$, and increase $D$ or $M$ or refine the gradient
step adaptively.  Although for these values $M$ the energy is very
inaccurate, the gradient is still reliable, and the method typically
converges after about $1000$ iteration steps.

\textit{Outlook.---}%
In this work, we have devised a class of states for which expectation
values can be computed using Monte Carlo sampling, and which therefore can
be used as a variational ansatz. The central idea is that $\langle
n\ket\psi$ can be computed efficiently, as it can be represented as a
product of easily describable terms. In particular, we considered
string-bond
states where each factor is a trace of a matrix product, defined on
strings which are distributed over the system. Due to the flexibility in
the layout of the strings, one can adapt the method to the geometry of the
underlying system, making it an interesting approach for problems in e.g.\
quantum chemistry or irregular systems.

The computation time scales as $D^3$ in the bond dimension ($D^2$ for open
boundaries), which improves over the $D^5$ ($D^3$) scaling of DMRG.
The reason is that the tensor network to be contracted has dimension $D$
rather than $D^2$. 
For the same reason, Monte Carlo sampling can also be
used to speed up the general PEPS method~\cite{peps-alg} from $D^{10}$ to
$D^6$.

Several extensions to the ideas presented in this paper are 
being investigated. Firstly, there are many choices for the factors
of $\langle n\ket\psi$ beyond strings, for instance other classes
developed for simulating many-body systems,  e.g.\ tree tensor
networks~\cite{ttn} which can be  arranged in a way reflecing the geometry
of the system. More generally, on small sets of spins (``blobs'') one
can allow for arbitrary states or for states parametrized e.g.\ by a
tensor network.  Note that in all these cases, the dependence
$\ket{\psi_A}$ on $A$ remains linear.  Finally, the string patterns (or
general partitionings) can even depend on the classical configuration
$\ket{n}$, corresponding to a dependence of the partitionings in
(\ref{eq:string-proj}) on the physical spin.

Beyond the simulation of ground state properties, string-bond states can also
be used to study time dependent phenomena or systems at finite
temperature, similar to DMRG and the general PEPS algorithm. Finally, the
ansatz can also be applied to fermionic systems, either by doing a
Jordan-Wigner transformation which does not affect the computability of
$\langle n\ket\psi$, or directly by using a fermionic ansatz for the
variational Monte Carlo method.

\textit{Acknowledgements.---}%
We would like to thank V.\ Murg and T.~Roscilde for helpful discussions
and for providing us with numerical data. This work has been supported by
the EU (COVAQIAL, SCALA), the German 
cluster of excellence project MAP, 
the DFG-Forschergruppe 635, and the
Elite Network of Bavaria (QCCC).

\textit{Note added.---}The fact that Monte Carlo can be used in tensor
network contraction has been proposed independently in~\cite{sandvik}.

\end{document}